\documentstyle[aps,preprint,prl]{revtex}

\draft

\begin{document}
\title{Surface barrier dominated transport in NbSe$_2$}
\author{Y. Paltiel, D. T. Fuchs, E. Zeldov, Y. N. Myasoedov, and H. Shtrikman}
\address{Department of Condensed Matter Physics, The Weizmann Institute of
Science, Rehovot 76100, Israel}
\author{M. L. Rappaport}
\address{Physics Services, The Weizmann Institute of Science, Rehovot 76100,
Israel}
\author{E. Y. Andrei}
\address{Department of Physics and Astronomy, Rutgers University, Piscataway,
New Jersey 08855}
\date{\today}
\maketitle

\begin{abstract}
Transport current distribution in clean 2$H$-NbSe$_{2}$ crystals is studied
by measuring the self-induced magnetic field across the sample. Below 
$T_c$ most of the current flows at the edges of the crystals due to strong
surface barriers, which are found to dominate the transport properties and
the resistive transition. The measured critical current is determined by the
critical current for vortex penetration through the surface barrier rather
than by bulk pinning.
\end{abstract}



\newpage

Vortices in type-II superconductors have to overcome surface and geometrical
barriers (SB) in order to exit or penetrate into the superconductor. The
effect of SB on the magnetic properties has been extensively studied
theoretically \cite{bl,clem,burlachkov,GB}. In recent years numerous studies
have shown that SB dominate the magnetization behavior in clean crystals of
high temperature superconductor (HTSC), in particular at elevated
temperatures \cite{chikumoto,konczy,indenbom,vander,EPL,morozov}. Since bulk
pinning is strongly reduced by thermal fluctuations, the relative importance
of the SB is expected to grow with temperature \cite{burla}. At low
temperatures, in contrast, and in low-$T_{c}$ superconductors in general,
bulk pinning is expected to be the main source of hysteretic magnetization.
Recent theoretical works \cite{bur,benkra} have suggested that SB can also
significantly modify the transport properties of superconductors. Current
distribution measurements in Bi$_{2}$Sr$_{2}$CaCu$_{2}$0$_{8}$ (BSCCO)
crystals \cite{nat,PRL} have revealed that over a wide range of temperatures
and fields, the transport current indeed flows predominantly at the edges of
the crystals where vortices enter and exit the superconductor. In this paper
we report that SB dominate the transport behavior also in the low-$T_{c}$
superconductor NbSe$_{2}$. SB should therefore be of significant importance
in a wide range of superconductors in affecting the transport as 
well as the magnetic properties.

High purity 2$H$-NbSe$_{2}$ crystals with $T_{c}$=7.2K were grown as
described previously \cite{lee}. Several crystals were cleaved and 
cut into a rectangular strip shape. The data presented here are for 
one of the crystals with dimensions of 1.46 mm($l$) $\times $ 0.35 
mm($w$) $\times $ 0.04 mm($d$). The features reported below were 
observed in all the investigated crystals. Four pads for electrical 
contacts of 0.1 $\times $ 0.1 mm$^{2}$ with 0.13 mm separation 
between pairs were prepared by Ag/Au evaporation. The crystal was 
mounted onto an array of 19 two-dimensional electron gas GaAs/AlGaAs 
Hall sensors 10 $\times $ 10 $\mu $m$^{2}$ each with a 10 $\mu $m 
separation. The inset to Fig. 1 shows a schematic top view of the 
experimental setup. A dc magnetic field $H_{dc}$ was applied 
perpendicular to the plane of the sensors and parallel to the c-axis 
of the crystal.  An ac transport current $I_{ac}$ in the range of 1 
to 30 mA at 65 Hz was applied, and the corresponding self-induced 
magnetic field $B_{ac}(x)$ across the crystal was measured by the 
Hall sensor array. The four-probe resistance of the sample was 
measured under the same conditions using a lock-in amplifier.

There are three main regimes for the flow of the transport current as
described in Ref. \cite{nat}: ($a$) Uniform current flow in the bulk of the
sample. In this case, applying the Biot-Savart law, the perpendicular
component of the self-induced field, as measured by the sensors, decreases
monotonically from the left edge to the right. ($b$) Surface barrier
dominated flow for which most of the current flows at the two edges of the
crystal in order to drive the vortices over the entry and exit SB. The
resulting self-induced field across the crystal is opposite in sign and
increases from left to right. ($c$) Vortex pinning which prevents vortex
motion and results in zero self-field inside the sample. The corresponding
transport current, in this case, has a characteristic Meissner distribution
with a reduced current in the center and enhanced at the edges \cite
{zel,brandt,norris}.

Figure 1a shows an example of the self-induced field $B_{ac}$ as a
function of temperature at $H_{dc}$=0.1T and $I_{ac}$=6mA. For clarity,
only the sensors 3 to 18, which are under the crystal, are shown. The
corresponding resistance is shown in Fig. 1b. The observed behavior 
is very similar to that reported for BSCCO crystals \cite{nat}. 
Above $T_{c}$, the current flows uniformly as expected, and 
$B_{ac}$ decreases monotonically from the left edge of the crystal 
(sensor 3) to the right edge (sensor 18). This behavior is seen more 
clearly in Fig. 2a, which shows $B_{ac}(x)$ profile for all the 
sensors at 7.6K. As the temperature is decreased, the SB sets in 
immediately below $T_{c}$ drawing an increasingly larger fraction of the
current to the edges. This gives rise to the observed temperature dependence
of $B_{ac}$ in Fig. 1a: a drop in $B_{ac}$ just below $T_{c}$,
followed by crossing of the curves and sign reversal at lower temperatures.
At the crossing point, half of the transport current flows across the bulk
and the other half at the two edges of the sample. At $T=T_{max}$ (see
Fig. 1a) most of the current flows at the crystal edges and the self-field
profile inside the crystal is completely inverted relative to the uniform
flow case, as shown in Fig. 2a for $T$=6.8K. Note that any bulk mechanism or
finite skin depth effect will result in $B_{ac}$ which is either zero
(perfect shielding) or positive in the left half of the sample (finite
shielding), but it cannot cause $B_{ac}$ to become negative (negative
permeability \cite{morozov}). The sign reversal of $B_{ac}$ is a unique
property of the SB. As the temperature is further decreased, the vortices
become immobile for $T<T_{d}$ due to the combination 
of the SB and bulk pinning, resulting in a vanishing $B_{ac}$ response
within the crystal as shown by the 4.9 K profile in Fig. 2a.

In Ref. \cite{nat} arrays of seven Hall sensors were used, which allowed
mapping of $B_{ac}$ typically over only half of the sample width. Here we
have extended the arrays to 19 sensors that provide more detailed 
$B_{ac}(x)$ over the entire sample width. This improvement has two significant
advantages. First, we can readily examine both edges of the sample
and evaluate the symmetry of the current distribution. The second major
advantage is that having 19 values of $B_{ac}$ field across the sample is
sufficient in order to directly invert the field distribution into the
current distribution. We represent the sample by 19 current filaments
located at about half of the crystal thickness and equally spaced across the
width. The currents in the filaments are obtained by inverting the 19 
$\times $ 19 matrix that transforms between the current and the field values
using the Biot-Savart law. Figure 2b shows the obtained $I(x)$
corresponding to the three field profiles in Fig. 2a. As expected, above 
$T_{c}$ the current flows uniformly across the crystal (7.6K profile).
However, below $T_{c}$ the current starts to accumulate at the edges due
to the strong SB. At $T=T_{max}$ practically all the current flows at
the two edges in a form of two $\delta $ functions with negligible current
in the bulk (6.8K profile). At low temperatures, 4.9K profile, the vortices
become immobile and the current is distributed in the corresponding
self-shielding form \cite{zel,brandt,norris} with a minimum in the center
and a rapid increase near the edges. Note that the actual current
distribution is continuous across the width and the thickness of the
crystal, and hence our derivation of the discreet current filaments in Fig.
2b is only an approximation. Yet, this simple analysis clearly visualizes
the underlying mechanisms, and in particular the main transition from a
uniform current flow above $T_{c}$ to SB dominated flow below $T_{c}$.
This finding shows that in clean NbSe$_{2}$ crystals the vortex flow rate 
is determined by the transmission probability through the SB rather than by
bulk vortex dynamics. An important experimental implication is that the
transport measurements in this case reflect the resistive properties of the
SB rather than the bulk properties.

Another aspect of the SB is its asymmetry with respect to vortex entry and
exit \cite{bl}. Vortex entry requires a larger force than vortex exit. As a
result a larger current flows at the vortex entry edge in order to maintain
the same vortex flow rate throughout the sample. The role of the edges is
interchanged as the direction of the ac current changes, with a larger
current flowing on the opposite edge of the crystal. This mechanism results
in a significant local second harmonic self-field signal, as shown 
in Fig. 3 for $H_{dc}$=0.5T and $I_{ac}$=6mA. Second harmonic is 
a unique feature of the SB due to its asymmetry with respect to the 
current direction. Bulk vortex dynamics, in contrast, results only 
in odd harmonics, since bulk I-V characteristics are antisymmetric 
with respect to the current. Figure 3 shows that the second harmonic 
signal, and hence the SB, set-in immediately below $T_{c}$ 
concurrently with the resistive drop. The narrow dip in the 
resistance at $T_{p}$ in Fig. 3 is the common peak effect in NbSe$_{2}$ 
\cite{bhatt,henderson,ghosh,DAnna}, which in our high purity crystals is
extremely narrow (see also Fig. 1a). Within our experimental resolution we
do not observe substantial changes in current distribution in this narrow
peak-effect region.

In contrast to BSCCO, the thermal activation of vortices in NbSe$_{2}$ is
weak due to much lower $T_{c}$ and lower anisotropy. We can therefore
simplify the description by analyzing the behavior in Fig. 3 in terms of
bulk and surface barrier critical currents, $I_{c}^{b}$ and 
$I_{c}^{s}$, respectively.  Let us first consider the case of finite 
bulk pinning with no SB, namely $I_{c}^{s}$=0. The bulk critical 
current $I_{c}^{b}$ is generally expected to increase with 
decreasing temperature, but as long as the applied current is larger 
than $I_{c}^{b}$($T$), the current should flow uniformly across the 
sample \cite{zel,brandt,norris}. As a result, $B_{ac}$ should be positive
like above $T_{c}$, and there should be no second harmonic signal. 
When $I_{c}^{b}(T)$ approaches the value of the applied current at 
some characteristic temperature $T_{max}$, vortices stop moving, and 
$B_{ac}$ should drop rapidly from the full positive value to zero as 
the temperature is further decreased. This scenario is inconsistent 
with the data in Fig. 3a: $B_{ac}$ is negative instead of
positive above $T_{max}$, and in addition, significant second harmonic is
present. Furthermore, below 
$T_{max}$, $B_{ac}$ does drop rapidly, but this drop occurs 
while $B_{ac}$ is negative rather than positive.

We now consider the opposite scenario of finite SB with no bulk pinning,
$I_{c}^{b}$=0.
Since $I_{c}^{s}$($T$) is related to the critical field $H_{c}$($T$), it
is expected to grow rather linearly below $T_{c}$. Therefore, upon cooling,
a progressively larger fraction of the applied current should be 
drawn towards the
edges. The corresponding $B_{ac}$ should drop approximately linearly 
from a full positive value
at $T_{c}$ down to a full negative value at a characteristic 
temperature at which $I_{c}^{s}$($T$) reaches $I_{ac}$, 
and as a result practically all the applied current flows at the 
edges. In this temperature interval a 
significant second harmonic signal should be also present, as 
explained above. The corresponding resistivity of the sample should 
decrease gradually below $T_{c}$ with a sharp drop towards zero at 
the same characteristic temperature. The data in Fig. 3 above 
$T_{max}$ is fully consistent with this scenario, thus allowing us 
to identify $T_{max}$ as the characteristic temperature at which 
$I_{ac}$ = $I_{c}^{s}$($T$).

Now let us examine the behavior below $T_{max}$ where 
$I_{ac}$ \mbox{$<$} $I_{c}^{s}$($T$) and no vortices can exit or 
penetrate into the sample (in absence of thermal activation).  Yet, 
if $I_{c}^{b}$=0 the vortices can move freely {\em inside} the 
sample and change their distribution according to $B_{ac}$ imposed 
by the full $I_{ac}$ flowing on the edges.  Since the edges have 
already absorbed all of $I_{ac}$, there should be no further change 
in the current distribution below $T_{max}$, and $B_{ac}$ should 
remain constant at its full negative value (as observed in BSCCO 
\cite {nat}). On the other hand, if a small bulk $I_{c}^{b}$  
is present in addition to $I_{c}^{s}$, $B$$_{ac}$ should drop 
sharply to zero below $T_{max}$ since vortices become immobile. 
$T_{max}$ in this case corresponds to the temperature at which 
$I_{ac}$ becomes equal to the total critical current $I_{c}(T)= 
I_{c}^{b} + I_{c}^{s}$. The pronounced drop of $B_{ac}$ at $T_{max}$ 
in Fig. 3a indicates therefore the existence of some finite bulk 
$I_{c}^{b}$.  The fact that $B_{ac}$ is practically fully inverted 
at $T_{max}$ shows, however, that $I_{c}^{b}$ $\ll $ 
$I_{c}^{s}$. Note also that the second harmonic signal should be 
absent below $T_{max}$, since no vortices penetrate through the SB, 
consistent with the data in Fig. 3a. The tail 
of $B_{ac}$ as well as the weak resistive tail below $T_{max}$ 
indicate, furthermore, that a small flux creep is present resulting 
in some finite vortex motion.  Note, that even in this weak creep 
regime $B_{ac}$ is inverted, showing that most of the current flows 
at the edges and vortex dynamics is governed by SB.  The vortices 
become fully immobile only at a lower temperature $T_{d}$, below 
which the creep stops within our resolution.

Finally, we demonstrate here that the above condition, that the total
critical current $I_{c}$ is determined mainly by the critical current of
the SB, holds over the entire investigated range of temperatures and fields.
At low temperatures and relatively low current, the vortices are immobile
and the current flows with the characteristic Meissner distribution
resulting in vanishing $B_{ac}$. In this case one concludes that the
applied current $I_{ac}$ is below the total critical current $I_{c}$ =
$I_{c}^{b}$ + $I_{c}^{s}$, but the values of the individual critical
currents cannot be determined. In order to gain this important information
one has to increase the applied current just slightly above the total
critical current. In this situation the applied current is precisely divided
between the bulk and the edges according to $I_{c}^{b}$ and $I_{c}^{s}$.
As a result the vortices are set in motion, and the corresponding 
$B_{ac}$ signal provides the information on the relative importance 
of $I_{c}^{b}$ and $I_{c}^{s}$. Such a measurement is presented in 
Fig. 4 which shows $B_{ac}$, as measured by sensor 3, at various 
$I_{ac}$ between 5 and 30 mA at 0.3T. The arrows indicate 
the temperatures $T_{max}$ at which $I_{ac}$ = $I_{c}$($T$) = 
$I_{c}^{b}$ + $I_{c}^{s}$ for the various applied currents as 
described above.  The corresponding $I_{c}$($T$) dependence is shown 
in the inset. We find that in all cases when $I_{ac}$ reaches 
$I_{c}$($T$) the corresponding $B _{ac}$ signal is fully inverted 
(as the 6.8K profile in Fig. 2a) indicating that practically all the 
current flows at the edges, and that $I_{c}^{b}$ $\ll $ $I_{c}^{s}$. 
Thus, in clean crystals of NbSe$_{2}$ the measured $I_{c}$ reflects 
mainly the critical current of the SB and not the bulk critical 
current.

In summary, the distribution of transport current in clean platelet crystals
of 2$H$-NbSe$_{2}$ was studied by measuring the self-induced magnetic field.
The use of extended arrays of Hall sensors allows for a direct inversion of
the self-field profile into the current distribution profile across the
crystal width. Below $T_{c}$ the current is found to flow
predominantly at the sample edges due to strong surface barriers. The SB
govern the apparent resistivity of the crystals. Furthermore, the measured
critical current is determined by the critical current of the SB barrier
rather than by bulk critical current.

We are grateful to A. Wold for the high purity NbSe$_{2}$ crystals, 
and to G. Jung and S. Bhattacharya for helpful discussions. This 
work was supported by grant no 97-00424 from the United 
States-Israel Binational Science Foundation (BSF), Jerusalem, 
Israel, and by the MINERVA Foundation, Munich, Germany.

\newpage 

\ FIGURE CAPTIONS

Fig. 1. (a) Transport current self-induced field $B_{ac}$ as measured by
sensors 3 to 18 across the width of a NbSe$_{2}$ crystal at $H_{dc}$ =
0.1T and $I_{ac}$ = 6mA. At the crossing point of the curves half of the
transport current flows in the bulk and half at the sample edges. At 
lower temperatures the inverted curves indicate that most of the 
current flows at the edges. Vortices are immobile below $T_{d}$. (b) 
Corresponding four probe resistance measurement. The resistance 
drops sharply at $T_{max}$ and shows a tail at lower temperature due 
to weak flux creep. Inset: schematic cross section of the sample 
mounted on an array of 19 Hall sensors.

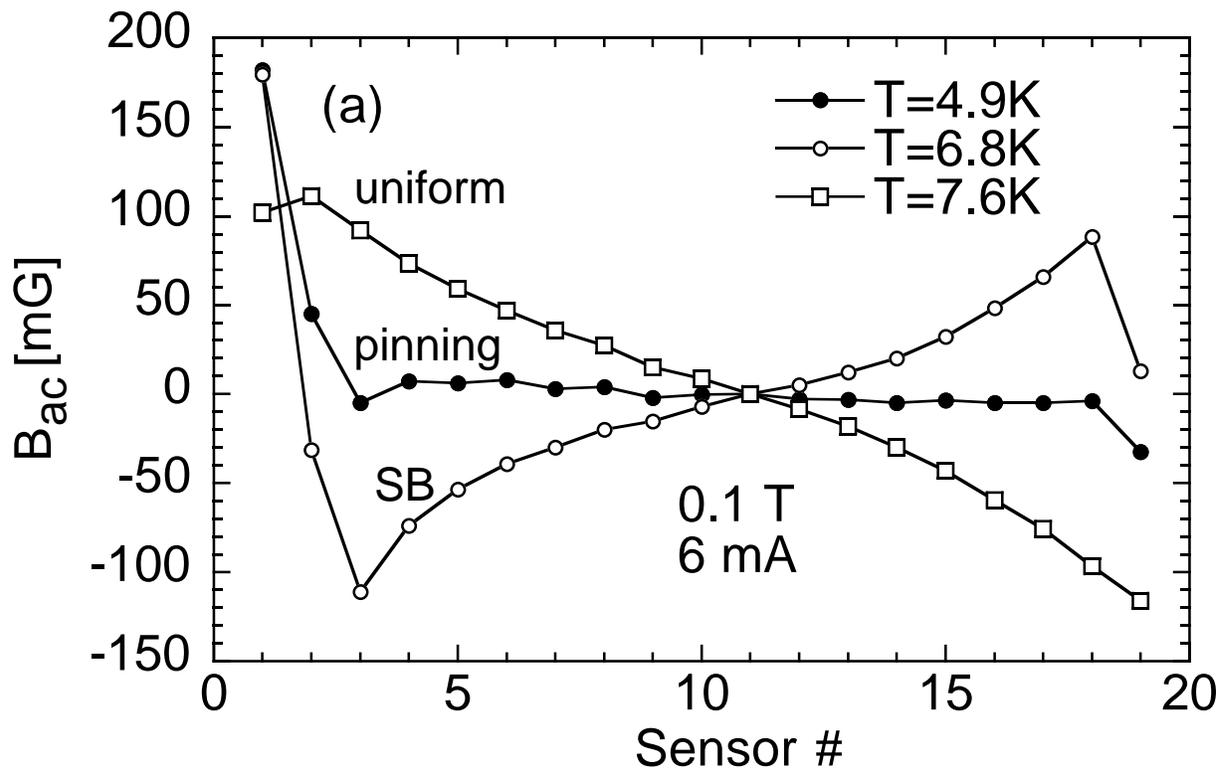
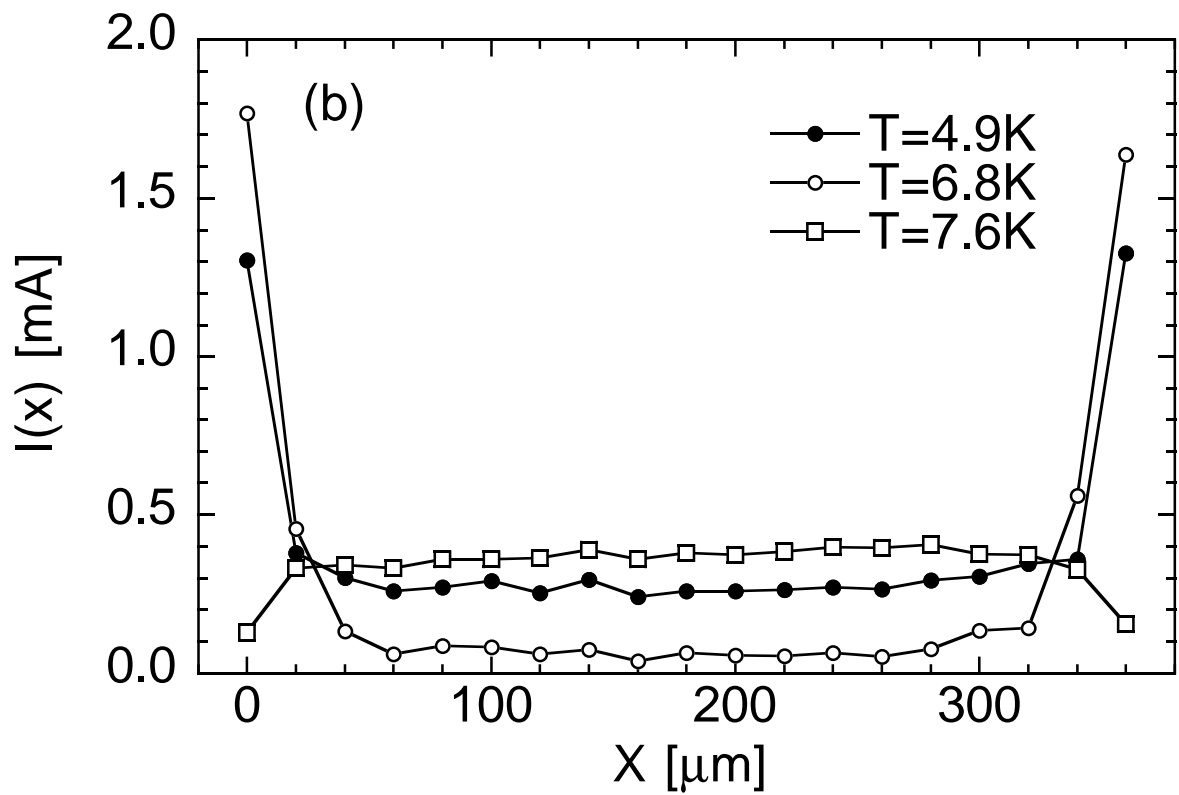
Fig. 2. (a) Three $B_{ac}(x)$ profiles at different temperatures taken
from Fig. 1a. (b) Transport current distributions obtained by inversion of
$B_{ac}$ in (a). Above $T_{c}$ the current flows uniformly across the
crystal (7.6K profile). At $T$ =$T_{max}$= 6.8K practically all the
current flows at the two edges of the sample due to strong SB. At
low temperatures the vortices are immobile and the current is distributed in
the corresponding self-shielding form (4.9K profile).

Fig. 3. First and second (multiplied by 3) harmonics of $B_{ac}$ as
measured by sensor 3 (a), and the corresponding resistance (b) at $H_{dc}$
= 0.5T and $I_{ac}$ = 6mA. The SB, and hence the second harmonic signal,
set in immediately below $T_{c}$. At $T_{max}$, $I_{ac}$ 
=$ I_{c}^{b}$ + $I_{c}^{s}$ and $B_{ac}$, the second harmonic 
signal, and the resistance drop sharply. The narrow dip in the 
resistance at $T_{p}$ is the peak effect in NbSe$_{2}$.

Fig. 4. $B_{ac}$ as measured by sensor 3 for various $I_{ac}$ between 5
and 30 mA at $H_{dc}$ = 0.3 T. As the current is increased vortices become
mobile at progressively lower temperatures. The arrows indicate the
temperatures at which the corresponding $I_{ac}$ equals $I_{c}$($T$).
The negative $B_{ac}$ values at these temperatures show that $I_{c}^{b}$ 
$\ll $ $I_{c}^{s}$ . Inset: temperature dependence of the total critical
current $I_{c}$ at $H_{dc}$ = 0.3T.


\end{document}